\documentclass[pra,twocolumn,showpacs,superscriptaddress,floatfix]{revtex4}
\font\titolo=cmbx12
\font\titolone=cmbx10 scaled\magstep 2%
\font\sc=cmcsc10\font\css=cmcsc8%
\font\ottorm=cmr8%
\font\msytw=msbm9 scaled\magstep1%
\font\msytww=msbm7 scaled\magstep1%
\font\msytwww=msbm5 scaled\magstep1%
\def\st{\scriptstyle}%
\def\dt{\displaystyle}%
\font\tenmib=cmmib10 \font\eightmib=cmmib8
\font\sevenmib=cmmib7\font\fivemib=cmmib5 
\font\ottoit=cmti8\font\fiveit=cmti5\font\sixit=cmti6
\font\fivei=cmmi5\font\sixi=cmmi6\font\ottoi=cmmi8
\font\ottorm=cmr8
\font\ottosy=cmsy8\font\sixsy=cmsy6\font\fivesy=cmsy5
\font\ottobf=cmbx8\font\sixbf=cmbx6\font\fivebf=cmbx5%
\font\ottocss=cmcsc8%

\def\ottopunti{\def\rm{\fam0\ottorm}\def\it{\fam6\ottoit}%
\def\bf{\fam7\ottobf}%
\textfont1=\ottoi\scriptfont1=\sixi\scriptscriptfont1=\fivei%
\textfont2=\ottosy\scriptfont2=\sixsy\scriptscriptfont2=\fivesy%
\textfont4=\ottocss\scriptfont4=\sc\scriptscriptfont4=\sc%
\textfont5=\eightmib\scriptfont5=\sevenmib\scriptscriptfont5=\fivemib%
\textfont6=\ottoit\scriptfont6=\sixit\scriptscriptfont6=\fiveit%
\textfont7=\ottobf\scriptfont7=\sixbf\scriptscriptfont7=\fivebf%
\setbox\strutbox=\hbox{\vrule height7pt depth2pt width0pt}%
\normalbaselineskip=9pt\rm}
\let\nota=\ottopunti%

\mathchardef\Ba   = "050B  
\mathchardef\Bb   = "050C  
\mathchardef\Bg   = "050D  
\mathchardef\Bd   = "050E  
\mathchardef\Be   = "0522  
\mathchardef\Bee  = "050F  
\mathchardef\Bz   = "0510  
\mathchardef\Bh   = "0511  
\mathchardef\Bthh = "0512  
\mathchardef\Bth  = "0523  
\mathchardef\Bi   = "0513  
\mathchardef\Bk   = "0514  
\mathchardef\Bl   = "0515  
\mathchardef\Bm   = "0516  
\mathchardef\Bn   = "0517  
\mathchardef\Bx   = "0518  
\mathchardef\Bom  = "0530  
\mathchardef\Bp   = "0519  
\mathchardef\Br   = "0525  
\mathchardef\Bro  = "051A  
\mathchardef\Bs   = "051B  
\mathchardef\Bsi  = "0526  
\mathchardef\Bt   = "051C  
\mathchardef\Bu   = "051D  
\mathchardef\Bf   = "0527  
\mathchardef\Bff  = "051E  
\mathchardef\Bch  = "051F  
\mathchardef\Bps  = "0520  
\mathchardef\Bo   = "0521  
\mathchardef\Bome = "0524  
\mathchardef\BG   = "0500  
\mathchardef\BD   = "0501  
\mathchardef\BTh  = "0502  
\mathchardef\BL   = "0503  
\mathchardef\BX   = "0504  
\mathchardef\BP   = "0505  
\mathchardef\BS   = "0506  
\mathchardef\BU   = "0507  
\mathchardef\BF   = "0508  
\mathchardef\BPs  = "0509  
\mathchardef\BO   = "050A  
\mathchardef\BDpr = "0540  
\mathchardef\Bstl = "053F  

\def\TTT{\hbox{\msytw T}} 

\def\RRR{\hbox{\msytw R}} 

\def\ZZZ{\hbox{\msytw Z}} \def\zzzz{\hbox{\msytww Z}}
\def\zzz{\hbox{\msytwww Z}}

\global\newcount\numsec\global\newcount\numapp
\global\newcount\numfor\global\newcount\numfig
\global\newcount\numsub
\def\veroparagrafo{\number\numsec}\def\veraformula{\number\numfor}
\def\veraappendice{\number\numapp}\def\verasub{\number\numsub}
\def\verafigura{\number\numfig}

\def\senondefinito#1{\expandafter\ifx\csname#1\endcsname\relax}

\def\SIA #1,#2,#3 {\senondefinito{#1#2}%
\expandafter\xdef\csname #1#2\endcsname{#3}\else
\write16{???? ma #1#2 e' gia' stato definito !!!!} \fi}

\def \Fe(#1)#2{\SIA fe,#1,#2 }
\def \Fp(#1)#2{\SIA fp,#1,#2 }
\def \Fg(#1)#2{\SIA fg,#1,#2 }

\def\Section(#1,#2){\advance\numsec by 1\numfor=1\numsub=1\numfig=1%
\SIA p,#1,{\veroparagrafo} %
\write15{\string\Fp (#1){\secc(#1)}}%
\hbox to \hsize{\titolo\hfill \number\numsec. #2\hfill%
\expandafter{\alato(sec. #1)}}\*}

\def\Appendix(#1,#2){\advance\numapp by 1\numfor=1\numsub=1\numfig=1%
\SIA p,#1,{A\veraappendice} %
\write15{\string\Fp (#1){\secc(#1)}}%
\hbox to \hsize{\titolo Appendix A\number\numapp. #2\hfill%
\expandafter{\alato(app. #1)}}%
\*%
}

\def\etichetta(#1){(\veroparagrafo.\veraformula)%
\SIA e,#1,(\veroparagrafo.\veraformula) %
\global\advance\numfor by 1%
\write15{\string\Fe (#1){\equ(#1)}}%
}

\def\etichettaa(#1){(A\veraappendice.\veraformula)%
\SIA e,#1,(A\veraappendice.\veraformula) %
\global\advance\numfor by 1%
\write15{\string\Fe (#1){\equ(#1)}}%
}

\def\getichetta(#1){\veroparagrafo.\verafigura%
\SIA g,#1,{\veroparagrafo.\verafigura} %
\global\advance\numfig by 1%
\write15{\string\Fg (#1){\graf(#1)}}%
}

\def\etichettap(#1){\veroparagrafo.\verasub%
\SIA p,#1,{\veroparagrafo.\verasub} %
\global\advance\numsub by 1%
\write15{\string\Fp (#1){\secc(#1)}}%
}

\def\Eq(#1){\eqno{\etichetta(#1)\alato(#1)}}
\def\eq(#1){\etichetta(#1)\alato(#1)}
\def\Eqa(#1){\eqno{\etichettaa(#1)\alato(#1)}}
\def\eqa(#1){\etichettaa(#1)\alato(#1)}
\def\eqg(#1){\getichetta(#1)\alato(fig. #1)}
\def\sub(#1){\0\palato(p. #1){\bf \etichettap(#1).}}
\def\asub(#1){\0\palato(p. #1){\bf \etichettapa(#1).}}
\def\apprif(#1){\senondefinito{e#1}%
\eqv(#1)\else\csname e#1\endcsname\fi}

\def\equv(#1){\senondefinito{fe#1}$\clubsuit$#1%
\write16{eq. #1 non e' (ancora) definita}%
\else\csname fe#1\endcsname\fi}
\def\grafv(#1){\senondefinito{fg#1}$\clubsuit$#1%
\write16{fig. #1 non e' (ancora) definito}%
\else\csname fg#1\endcsname\fi}
\def\secv(#1){\senondefinito{fp#1}$\clubsuit$#1%
\write16{par. #1 non e' (ancora) definito}%
\else\csname fp#1\endcsname\fi}

\def\eqo{{\global\advance\numfor by 1}}
\def\equ(#1){\senondefinito{e#1}\equv(#1)\else\csname e#1\endcsname\fi}
\def\graf(#1){\senondefinito{g#1}\grafv(#1)\else\csname g#1\endcsname\fi}
\def\figura(#1){{\css Figura} \getichetta(#1)}
\def\secc(#1){\senondefinito{p#1}\secv(#1)\else\csname p#1\endcsname\fi}
\def\sec(#1){{\secc(#1)}}
\def\refe(#1){{[\secc(#1)]}}

\def\BOZZA{
\def\alato(##1){\rlap{\kern-\hsize\kern-.5truecm{$\scriptstyle##1$}}}
\def\palato(##1){\rlap{\kern-.5truecm{$\scriptstyle##1$}}}
}

\def\alato(#1){}
\def\galato(#1){}
\def\palato(#1){}

{\count255=\time\divide\count255 by 60 \xdef\hourmin{\number\count255}
        \multiply\count255 by-60\advance\count255 by\time
   \xdef\hourmin{\hourmin:\ifnum\count255<10 0\fi\the\count255}}

\def\oramin{\hourmin }

\def\data{\number\day/\ifcase\month\or gennaio \or febbraio \or marzo \or
aprile \or maggio \or giugno \or luglio \or agosto \or settembre
\or ottobre \or novembre \or dicembre \fi/\number\year;\ \oramin}
\setbox200\hbox{$\scriptscriptstyle \data $}

\let\a=\alpha \let\b=\beta  \let\g=\gamma  \let\d=\delta  \let\e=\varepsilon
\let\z=\zeta     \let\th=\theta \let\k=\kappa  \let\l=\lambda
\let\m=\mu    \let\n=\nu    \let\x=\xi     \let\p=\pi     \let\r=\rho
\let\s=\sigma \let\t=\tau   \let\f=\varphi 
  \let\ps=\psi  \let\o=\omega  
\let\G=\Gamma \let\D=\Delta \let\Th=\Theta  
        
\let\O=\Omega 

\def\CC{{\cal C}}
\def\EE{{\cal E}}\def\FF{{\cal F}}
\def\II{{\cal I}}\def\KK{{\cal K}}\def\LL{{\cal L}}
\def\PP{{\cal P}}
\def\ScS{{\cal S}}\def\TT{{\cal T}}

\let\ig=\int
\let\io=\infty
\let\==\equiv
\let\0=\noindent

\def\\{\hfill\break}
\def\lis#1{\overline#1}

\def\*{\vskip3mm} 
\def\ie{{\it i.e. }}

\let\dpr=\partial
 
\def\defi{\,{\buildrel {\rm def}\over=}\,}

\def\V#1{{\bf#1}}
\def\media#1{{\langle#1\rangle}}
\def\fra#1#2{{#1\over#2}}
\def\crcl{\,\raise.5mm\hbox{$\st\rm o$}\,}%
\def\otto{\,{\kern-1.truept\leftarrow\kern-5.truept\to\kern-1.truept}\,}
\def\tende#1{\,\vtop{\ialign{##\crcr\rightarrowfill\crcr
 \noalign{\kern-1pt\nointerlineskip} \hskip3.pt${\scriptstyle
 #1}$\hskip3.pt\crcr}}\,}
\def\kap{\mathop\cap}%

\newdimen\xshift \newdimen\xwidth \newdimen\yshift \newdimen\ywidth

\def\ins#1#2#3{\vbox to0pt{\kern-#2\hbox{\kern#1 #3}\vss}\nointerlineskip}

\def\eqfig#1#2#3#4#5{
\par\xwidth=#1 \xshift=\hsize \advance\xshift
by-\xwidth \divide\xshift by 2
\yshift=#2 \divide\yshift by 2%
{\hglue\xshift \vbox to #2{\vfil
#3 \includegraphics{#4.ps}
}\hfill\raise\yshift\hbox{#5}}}

\def\8{\write12}

\input #1.eqx \closein13\fi
\input \jobname.eqx \closein14 \fi
\immediate\openout15=\jobname.eqx

\newcommand\revtex{{\nota R\kern-0.4mm\lower0.5mm\hbox{E}%
\kern-0.4mm V\kern-0.3mm%
\lower0.5mm\hbox{T}\kern-0.4mm E\kern-.3mm \lower0.5mm\hbox{X}}}
\newcommand\fancyhdr{{{\nota F\kern-1mm\lower0.5mm\hbox{A}%
\kern-0.6mm N\kern-0.5mm%
\lower0.5mm\hbox{C}\kern-0.5mm Y\kern-.5mm\lower0.5mm\hbox{H}%
\kern-0.3mm\hbox{D}\kern-0.45mm\lower0.5mm\hbox{R}}}}
\newcommand\EqaligN{{{\nota E\kern-0.3mm\lower0.5mm\hbox{Q}%
\kern-0.4mm A\kern-0.4mm%
\lower0.5mm\hbox{L}\kern-0.20mm I\kern-.30mm\lower0.4mm\hbox{G}%
\kern-.35mm\hbox{N}\kern-0.42mm\lower0.5mm\hbox{N}\kern-.35mm\hbox{O}}}}

\usepackage{eqalignno}%
\usepackage{fancyhdr}%
{%
}%
\begin{document}

\renewcommand{\headrulewidth}{0pt}
\rhead{\thepage}
\let\headline=\lhead
\lfoot{\nota\it\today}

\centerline{\titolone A fluctuation theorem in}
\vskip1mm
\centerline{\titolone a random environment}
\*

\centerline{\it F.Bonetto, G.Gallavotti, G.Gentile}
\*
\centerline{\nota\it Mathematics, Georgia Tech., Atlanta}
\centerline{\nota\it I.N.F.N. Roma 1, Fisica Roma1}
\centerline{\nota\it Matematica, Universit\`a di Roma 3}

\* \0{\it Abstract:\rm\ A simple class of chaotic systems in a random
  environment is considered and the fluctuation theorem is extended
  under the assumption of reversibility.}  \*

\*\*
\Section(1,Random chaos)
\lhead{\nota\it 1. Random chaos}

\0A chaotic system in random environment is defined by 
\*

\0(1) an {\it environment} which can be in states $\Bo$ belonging 
to a space $\O$ and varying in time; 
the time evolution being given by a map $\t$ on $\O$;
\\
(2) a $\t$--invariant probability distribution $P$ on $\O$ describing
the statistical properties of the environment evolution;
\\
(3) a family of maps $x\to f_\Bo(x)$ of
a manifold $\FF$ ({\it phase space}) into itself.

\*

The space of the random events $\O$ will be supposed a space of
sequences of finitely many symbols $\Bo=\{\o_j\}_{j=-\io}^\io$,
for instance a sequence of spins $\o_{j}=\pm1$, with the usual metric
$d(\Bo,\Bo')=\sum_j|\o_j-\o'_j| 2^{-|j|}$, and with $\t$ the shift to
the left: $(\t\Bo)_{j}=\o_{j+1}$. A ``reflection'' operator will
be defined on $\O$ as $\Bo\otto\Bo^T$, with $(\Bo^T)_j=\o_{-j}$.
The probability distribution $P$ will be a $\t$--invariant and
reflection--invariant mixing process on $\O$, for instance a Bernoulli
shift or a Markov process or a general Gibbs distribution on $\O$.
Reflection--invariant means that $P(E)\=P(E^T)$ for every Borel set $E$, 
where $E^T$ is the image of $E$ under the reflection $\Bo\otto\Bo^T$.
The manifold $\FF$ will be a torus $\TTT^m$ and $f_\Bo$ will be supposed
to be close to a map $f$ which is a ``linear hyperbolic torsion'' of $\TTT^m$,
independent of the environment, \ie close to a map which is defined by the
action of a matrix with integer entries and determinant $1$ on the torus. 

For instance the point $x\in\FF$ can be visualized as a pair of clock
arms $x=\Bf=(\f_1,\f_2)$ frantically moving as
$$\Bf\to S_0(\Bf)=(\f_1+\f_2,\f_1).\Eq(e1.1)$$ 
The map is a ``linear hyperbolic torsion'' of $\TTT^2$ and is a paradigmatic
example of Anosov maps.

The restriction on the choice of $f$, however, is not dictated by a
simplicity requirement alone. The present technique is very special 
for the cases that we shall consider. However it should be possible 
to extend them to a more general situation using the ideas in \cite{BFG04}. 

The actual $f_\Bo$ will be small perturbations of $f$. For simplicity, 
for a small $\e$, we will take it of the form
$$x\to f_\Bo(x)=f(x)+\e \o_0 \ps(x) , \Eq(e1.2)$$
where $\o_0\in\{-1,1\}$ and $\ps$ is a trigonometric polynomial
periodic on $\TTT^m$. Therefore we are led to consider the dynamical
system $(\O\times\FF,\ScS_\e)$, with
$$\ScS_\e(\Bo,x)=(\t\,\Bo,\,f_{\Bo}(x)) . \Eq(e1.3)$$
In other words at every instant $t$ the next coin is flipped (or the
next spin state is observed) and the system point $x_t$ is moved by
$f(x_t)+\e\o_t\ps(x_t)$.  We shall denote $\V x=(\Bo,x)$ the points in
$\O\times\FF$.

According to standard terminology if $\m_0$ denotes the normalized
volume measure on $\FF$ (``Liouville measure'') the system will be
said to possess a well defined statistics if, for almost all
$(\Bo,x)\in \O\times\FF$ in the $P\times\m_0$-distribution sense, the
limit
$$\lim_{T\to\io}\fra1T\sum_{j=0}^{T-1} F( f^{*j}_\Bo(x))=\ig_\FF
\m({\rm d}y) F(y) \Eq(e1.4)$$ 
exists for all continuous ``observables'' $F$ on $\FF$ (hence for many
others), and is independent of $\V x=(\Bo,x)$, thus defining a
probability distribution $\m$ which will be called the {\it
statistics} of the motion or the SRB distribution; here $f^{*j}_\Bo(x),
\,j\in \ZZZ$ is defined as the composition of $j$
maps $f_{\t^{j-1}\Bo}\crcl \ldots \crcl f_{\Bo}$ if $j>0$ and
$f_{\t^{j}\Bo}^{-1}\crcl \ldots \crcl f^{-1}_{\t^{-1}\Bo}$ if $j<0$,
and $f^0_\Bo=1$. More ambitiously one could look for an
$\ScS_\e$-invariant distribution of the form
$$\m_{\rm srb}({\rm d}\Bo\,{\rm d}x)=
P({\rm d}\Bo)\m_{\Bo}({\rm d}x) , \Eq(e1.5)$$
with $\lim_{T\to\io}\fra1T\sum_{j=0}^{T-1} F(\ScS^{j}_\e(\Bo,x))$
existing for all continuous observables $F$ on $\O\times\FF$, apart
from a $0$-probability set with respect to $P\times\m_0$, and having
the form $\ig_{\BO\times\FF} F(\Bo',y) \m_{\rm srb}({\rm d}\Bo' {\rm d}y)
\defi \media{F}_{\rm srb}$; when existing, $\m_{\rm srb}$ is also
called the SRB distribution and it is related to \equ(e1.4)
by $\m({\rm d}x)=\ig_\O \m_{\Bo}({\rm d}x)\,P({\rm d}\Bo)$.

Systems more general than the above have been considered in
\cite{Ru97a} where
$$\s_\Bo(x)=-\log \Big|\det \fra{\dpr f_\Bo(x)}{\dpr x}\Big|\Eq(e1.6)$$
is introduced and called {\it entropy production} rate: furthermore it
is proved that $\s_+\defi \media{\s}_{\rm srb}\ge0$, see \equ(e1.5).
Here we shall mainly consider the case $\s_+>0$, which is the generic case.

Particular attention will be given to {\it reversible} systems: namely
systems for which there is a smooth map $\II: \II(\Bo,x)
=(\Bo'(\Bo),x'(\Bo,x))$, with $\Bo'(\Bo)\defi \Bo^T$, such that
$\II^2=\pm1$ and
$$\II\crcl\ScS^k_\e=\ScS^{-k}_\e\crcl \II\Eq(e1.7)$$ 
for some integer $k$.

Examples of such systems are rather simple to give in the non--random
case: however in the random cases (like the ones considered here) the
very existence of $\II$ is not obvious.

In this paper we follow the general ideas of the ``cluster expansion''
techniques which were introduced in \cite{BK95,JM96,JP98,BK96,BK97}
but developed following the tree expansion method of
\cite{BFG04}, see also \cite{GBG04}.

\*\*
\Section(2,Fluctuation theorem)

\0In the non--random case, \ie if $f_\Bo=f$ is independent of $\Bo$, and
\lhead{\nota\it 2. Fluctuation theorem}
for $f$ a general Anosov map it has been shown that a time reversal
symmetry can be translated into certain relations between the
probabilities of the ``large fluctuations'' of the time averages of
the dimensionless observable $\s_\Bo(x)/\s_+$. Namely consider
the observable
$$p(x)=p(\V x)\defi
\fra1T\sum_{j=0}^{T-1} \fra{\s_{\t^j\Bo}(f_\Bo^{*j}x)}{\s_+} , \Eq(e2.1)$$
and call $\p_T(p\in\D)$ the probability that it takes value in an
interval $\D$ evaluated in the stationary SRB distribution (which for
Anosov maps exists).  Then $\p_T(p\in\D)=e^{T \z(p)+O(1)}$ where
$\z(p)$ is an analytic function of $p$ defined in the interval
$(-p^*,p^*)$, for some $p^*>1$. The time reversal symmetry implies 
$$\z(-p)=\z(p)-p\,\s_+,\qquad |p|<p^* , \Eq(e2.2)$$
which is called the {\it fluctuation theorem}, and was proven in
\cite{GC95,GC95b,Ga95b}.

The above statements concern non--random maps (\ie $f_\Bo=f$ is $\Bo$
independent).  This note is dedicated to show that \equ(e2.2) also
holds for the random noise cases introduced in \equ(e1.3) with $f_\Bo$
given by \equ(e1.2) and $f=S_{0}$, as given for instance by \equ(e1.1),
when a time reversal symmetry, \equ(e1.7), $\II(\Bo,x)=(\Bo'(\Bo),
x'(\Bo,x)$ holds with $\Bo'(\Bo)\defi\Bo^T= \{\o_{-j}\}_{j=-\io}^\io$
and $P$ is invariant for the map $\Bo\otto\Bo^T$.

A proof can be achieved by trying to fall back on the already existing
proofs of extensions of the fluctuation theorem to stochastic
processes; the first one in \cite{Ku98} has been followed and widely
extended in \cite{LS99,Ma99}.

In the first examples mentioned at the beginning of Sec.\sec(1),
in which the $\Bo$'s are a sequence of $\pm1$ with independent
distribution or with finite range coupling,
it is clear that the process $\m({\rm d}\Bo\,{\rm d}x)$ is a Markov
process and the analysis in Sec.2 of \cite{LS99} can in principle be
applied.  In the present case the phase space is a continuum but the
results in the quoted paper nevertheless apply, at least if one is
willing to consider formally ratios of delta functions and interpret
them as suitable Jacobian determinants.

To discuss the example of a general Gibbs distribution,
in which the distribution of the $\Bo$ is far from Markovian,
along the lines of \cite{LS99} is harder. But 
one can consider the more general approach in \cite{Ma99}. However
in this case one has to prove that the SRB distribution is a ``space-time
 Gibbs state'', which is the basic object studied in \cite{Ma99}.

{\it In all cases an explicit determination} of the stationary state
seems {\it necessary} in order to check the assumptions and to define
and compute the quantity $e(\l)$ of the quoted references. In other
words the existing literature provides strong arguments (particularly
in the Markovian cases) for the validity of a relation like
\equ(e2.1): but some work to check it remains to be done, on a case by
case basis, and a substantial further work is necessary if one wishes
to compute the function $\z(p)$ or at least some of its properties.

We restrict attention to the (very special) cases

(1) $\ScS_\e$ defined by \equ(e1.2), \equ(e1.3), \equ(e1.1) and

(2) $\ps(x)$ trigonometric polynomial.

We shall show that the technique of \cite{BKL03} can be applied to
obtain, for $\e$ small enough, the results: \*

\0(i) {\it the SRB distribution $\m$ exists and is unique;} 
\\ 
(ii) {\it it is a space-time Gibbs distribution;}
\\
(iii) {\it $\s_+>0$ at least for $\e\neq 0$ small enough;}
\\
(iv) {\it under the extra assumption of existence of a time reversal
symmetry for $\ScS$, the large deviation rate $\z(p)$ for $p(x)$,
see \equ(e2.1), satisfies the symmetry relation \equ(e2.2), \ie a
fluctuation theorem holds.}  \*

\0{\it Remarks:} (i) The extension of the results to perturbations of
torsions of tori of arbitrary dimension
is not treated because it is identical to the one we present here but
the notational burden hides the simplicity of the ideas.  
\\ 
(2) The apparent restriction to perturbations of Anosov maps generated
by a linear torsion of a torus is due to the use made of the flatness
and parallelism properties of the stable (respectively unstable)
planes for such torsions. However by applying the methods in
\cite{BFG04} the results can be extended to the general
case in which perturbations of generic analytic Anosov maps of tori of
arbitrary dimension are considered, see comments in Sec.\sec(8).

\*\*
\Section(3,Decoupling and shadowing)
\lhead{\nota\it 3. Decoupling and shadowing}

\0The key for fulfilling the program set in Sec.\sec(2)  is to show
that there is a map $\V x'\=(\Bo',x')\otto\V x\=(\Bo,x)$, 
that we call $H_\e$, of the form
$$(\Bo',x')=H_\e(\Bo,x)\defi(\Bo,x+h_\Bo(x)) , \Eq(e3.1)$$
with $h_{\Bo}(x)$ analytic in $\e$ (but not in $x$, in general) such that
$$\ScS_\e\crcl H_\e=H_\e\crcl\ScS_0 ,\Eq(e3.2)$$
where $\ScS_0(\Bo,x)=(\t\Bo,S_0x)$.
This means that there is a change of variables turning the perturbed
map into the unperturbed one, \ie the perturbed map $\ScS_\e$ can be
``conjugated'' to the unperturbed one $\ScS_0$. The relation \equ(e3.2)
becomes, see \equ(e1.2),
$$S_0 h_\Bo(x)-h_{\t\Bo}(S_0x)=-\e\ps_\Bo(x+ h_\Bo(x)) , \Eq(e3.3)$$
where $S_0$ is a $2\times2$ matrix in our case.

We look for a solution which is analytic in $\e$: $h_\Bo(x)=\e
h^{(1)}_\Bo(x)+\e^2 h_\Bo^{(2)}(x)+\cdots$, with $h^{(k)}_\Bo(x)$ 
$\e$--independent functions.  For instance the equation for the first
order is
$$ S_0h_\Bo^{(1)}(x)-h_{\t\Bo}^{(1)}(S_0 x)=- \ps_\Bo(x). \Eq(e3.4)$$
Call $v_{\pm}$ the two normalized eigenvectors of $S$
relative to the eigenvalues $\l_\pm$ and let $\l$ the
inverse of the largest one ($\l=\fra12(\sqrt 5-1)$), so that
$\l_+=\l^{-1},\l_-=-\l$: $\l<1$.

The functions $\ps_\Bo, h_\Bo$ can be split into two components
along the vectors $v_\pm$:
$$ \eqalign{
\ps_\Bo(x)= \ps_{\Bo,+}(x)v_++ \ps_{\Bo,-}(x)v_- , \cr
h_\Bo(x)= h_{\Bo,+}(x)v_++ h_{\Bo,-}(x)v_- , \cr
}\Eq(e3.5)$$
and the equation \equ(e3.5) for $h_{\Bo,\pm}^{(1)}$ gives
$$ \eqalign{
\l_+ &h_{\Bo,+}^{(1)}(x)-h_{\t\Bo,+}^{(1)}(S_0x) =\,\ps_{\Bo,+}(x) , \cr
\l_- & h_{\Bo,-}^{(1)}(x)-h_{\t\Bo,-}^{(1)}(S_0x)=\,\ps_{\Bo,-}(x) , \cr
}\Eq(e3.6)$$
which can be solved by simply setting
$$ h_{\Bo,\a}^{(1)}(x) = -\sum_{p\in \zzzz_\a} \a \l_\a^{-|p+1|\a}
\ps_{\t^{p}\Bo,\a}(S^p_0x) , \Eq(e3.7)$$
where $\a=\pm$, $\ZZZ_+=[0,\io)\kap \ZZZ$, $\ZZZ_-=(-\io,0)\kap \ZZZ$
and the inequality $\l<1$ ensures convergence. 

Hence the equations for $h^{(k)}_{\pm}$ become
$$ \eqalign{
&h^{(k)}_{\Bo,\a}(x)= - \sum_{s=0}^\io\fra1{s!}
\sum_{k_1+\cdots+k_s=k-1,\,k_i\ge0 \atop \a_1,\ldots,\a_s=\pm}\cdot\cr
&\cdot
\sum_{p\in \zzzz_\a} \a\l_\a^{-|p+1|\a} \Big(\prod_{j=1}^s
(v_{\a_j}\cdot\dpr_{x})
\Big) \ps_{\t^{p}\Bo,\a}(S_0^px)\cdot
\cr
&\cdot \Big(\prod_{j=1}^s
h^{(k_j)}_{\t^{p}\Bo,\a_j}(S_0^{p}x)\Big) . \cr}
\Eq(e3.8)$$
This can be written via a graphical representation as in Fig.1,
where the ``graph element'' on the l.h.s.
represents $h_{\Bo,\a}^{(k)}(x)$. Representing again,
in the same way, the graph elements that appear on the r.h.s. one
obtains an expression for $h^{(k)}_{\Bo,\a}(x)$ in terms of {\it trees},
oriented ``toward the root''.

\*
\eqfig{168.00000pt}{76.800003pt}{
\ins{-6.40000pt}{60pt}{$\a$}
\ins{13.60000pt}{63pt}{$(k)$}
\ins{24.00000pt}{58pt}{$\ =\kern-10pt
{\dt\sum_{s>0 , p \in \zzz_{\a} 
\atop k_{1}+\ldots+k_{s}=k-1}}\kern-7pt\fra1{s!}$}
\ins{95.20000pt}{60pt}{$\a,p$}
\ins{121.59999pt}{75pt}{$\a_1$}
\ins{121.59999pt}{35.200001pt}{$\a_s$}
\ins{147.19999pt}{28.800001pt}{$(k_{s})$}
\ins{147.19999pt}{41.600002pt}{$(k_{s-1})$}
\ins{147.19999pt}{70.400002pt}{$(k_2)$}
\ins{147.19999pt}{83.200005pt}{$(k_1)$}
}{fig1}{}
\*
\vglue-30pt
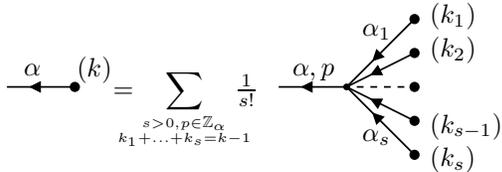
Fig.1 {\nota Graphical interpretation of \equ(e3.8)
for $k> 1$.}
\*

A tree $\th$ with $k$ nodes will carry on the branches
$\ell$ a pair of labels $\a_{\ell},p_{\ell}$,
with $p_{\ell}\in \ZZZ$ and $\a_{\ell}\in\{-,+\}$,
and on the nodes $v$ a pair of labels (not shown in Fig.1)
$\a_{v},p_{v}$, with $\a_{v}=\a_{\ell_{v}}$ and
$p_{v}\in \ZZZ_{\a_{v}}$ such that
$$ p(v)\=p_{\ell_{v}}=\sum_{w \succeq v} p_v , \Eq(e3.9)$$
where the sum is over the nodes following $v$ (\ie over the nodes
along the path connecting $v$ to the root), $\ell_{v}$ denotes the
branch exiting from the node $v$, and to each tree we shall
assign, given $\Bo$, a {\it value} given by
$$\eqalign{
{\rm Val}(\th)=&\prod_{v\in V(\th)}
\fra{-\a_v}{s_v!} \l_{\a_v}^{-|p_v+1|\a_v}\cdot\cr
&\cdot 
\Big( \prod_{j=1}^{s_v} \dpr_{\a_{v_j}} \Big)
\ps_{\t^{p(v)}\Bo,\a_v}(S_0^{p(v)}x) ,\cr} \Eq(e3.10)$$
where $\dpr_\a\defi v_\a\cdot\V\dpr_x$, $V(\th)$ is the set of nodes
in $\th$, the nodes $v_1,\ldots,v_{s_v}$ are the $s_v$ nodes preceding
$v$ (if $v$ is a top node then the derivatives are simply missing).
If $\Th_{k,\a}$ denotes the set of all trees with $k$ nodes and with
label $\a$ associated with the root line, then one has
$$ h_{\Bo,\a}(x)= \sum_{k=1}^{\io} \e^{k} \sum_{\th\in\Th_{k,\a}}
{\rm Val}(\th) , \Eq(e3.11)$$
and the ``only'' problem left is to estimate the radius of convergence
of the above formal power series. For this purpose it is convenient to
study the Fourier transform of the function $h_{\Bo,\a}(x)$. This is
easily done graphically because it is enough to attach a label
$\Bn_v\in \ZZZ^2$ to each node and define the {\it momentum} that
``flows'' on the tree branch $\ell_{v}$ as $ \Bn_{\ell_{v}} \defi
\sum_{w \preceq v} \Bn_w $.

Then \equ(e3.11) becomes
$$ h_{\Bo,\a}(x) = \sum_{k=1}^{\io} \e^{k}
\sum_{\Bn\in \zzzz^{2}} e^{i\Bn\cdot x} \, h^{(k)}_{\Bo,\a,\Bn} ,
\Eq(e3.12) $$
with
$$ \eqalign{
&h^{(k)}_{\Bo,\a,\Bn}=\sum_{\th\in\Th_{k,\Bn,\a}} \cdot\cr&\cdot
\Big( \prod_{v\in V(\th)} \fra{-\a_v}{s_v!}
\l_{\a_v}^{-|p_v+1|\a_v} \ps_{\t^{p(v)}\Bo,\a_v,S_0^{-p(v)}\Bn_v} \Big)
\cdot \cr\cdot
& \prod_{v\in V(\th)\atop v'\not=v_0}\big(
i S_0^{-p(v')}\Bn_{v'}\cdot{ v_{\a_v}} \big) , \cr} \Eq(e3.13)$$
where $\Th_{k,\Bn,\a}$ denotes the set of all trees with $k$ nodes and
with labels $\Bn$ and $\a$ associated with the root line,
and $v_0$ is the node immediatley preceding the root.

Since we are assuming that $\V \ps_\Bo$ is a trigonometric polynomial
of degree $N$ and with uniformly bounded coefficients as $\Bo$
varies we can define $F=\max_\Bn |\ps_{\Bo,\Bn}|$. Then, to study
the regularity of the conjugation, we shall estimate
$\sum_{\Bn} |\Bn|^\b |h^{(k)}_{\a,\Bn}|$ for $\b\ge 0$.
The only problem is given by the presence of the factor $|\Bn|^\b$.
In fact consider first the case $\b=0$:
there are only $(2N+1)^2<(3N)^2$ possible choices for each
$\Bn_v$, given $p(v)$, such that $|S_0^{-p(v)}\Bn_v|\le N$.
Hence fixed $\th$, $\{\a_{v}\}_{v\in V(\th)}$
and $\{p_v\}_{v\in V(\th)}$ the remaining sum of products
in \equ(e3.13) is bounded by (if $\l\=\l_+^{-1}\=-\l_-$)
$$ (3N)^{2k} N^{k} F^k \prod_{v \in V(\th)}
\fra{\l^{|p_v+1|}}{s_v!} , \Eq(e3.14)$$
having bounded by $N^k$ the last product in \equ(e3.13).  The sum over
the $\{p_v\}_{v\in V(\th)}$ is a multiple
geometric series bounded by $(2/(1-\l))^k$.

The combinatorial factors arising from the Taylor expansion leading to
\equ(e3.8) (see the $s!^{-1}$ factors in \equ(e3.8) or the
$s_{v}!^{-1}$ factors in \equ(e3.14))
compensate the large number of trees so that the factor $\prod_v
(1/s_v!)$ becomes, after summing over all the trees, simply bounded by
$2^{3k}$, ($2^{2k}$ due to a (poor) estimate of the number of trees,
after appropriate combinatorial considerations, and $2^{k}$ due to the
sum over the labels $\a_{v}=\pm$).

Therefore for $\b=0$ the conjugating function $H_\e$ exists and it is
uniformly continuous and uniformly bounded with a uniformly summable
Fourier transform, for $\e$ (even complex) in 
$|\e|<\e_0(0)\defi (3N)^{-3}F^{-1}2^{-4}(1-\l)$,
where an extra factor $2^{-1}$ has been inserted in order
to obtain uniform bounds.

Taking $\b>0$ requires estimating $|\Bn|^\b$: we bound it by
$\sum_v|\Bn_v|^\b$. Then from $|S_0^{-p(v)}\Bn_v| \le N$ (otherwise
the value vanishes), we infer that $|\Bn_v|\le \l^{-|p(v)|} BN$, where
$B\ge1$ is a suitable constant.  The sum $\sum_v|\Bn_v|^\b$ is over
$k$ terms which can be estimated separately so that we can write
$\sum_v|\Bn_v|^\b\le k |\Bn_{\lis v}|^\b$ where $|\Bn_{\lis
v}|=\max_{v}|\Bn_{v}|$.  This can be taken into account by multiplying
\equ(e3.14) by an extra factor $(B N)^\b\l^{-\b|p(\lis v)|} \le B N
\l^{-\b\sum_v |p_v|}$. Therefore if $\b<1$ the $\sum_{\Bn} |\Bn|^\b
|h^{(k)}_{\a,\Bn}|$ is bounded as in the case $\b=0$ but the result 
is modified into
$$ \e_0(\b) = (3N)^{-3} F^{-1}(1-\l^{1-\b}) 2^{-5} . \Eq(e3.15)$$
This shows that $H_{\e}$ is analytic in $\e$ in the disk
with radius $\e_0(\b)$. Furthermore, since in \equ(e3.15) we
inserted (for simplicity) an extra factor $2^{-1}$ in excess of the result
obtained by the procedure described, the H\"older modulus is also
uniformly bounded by a suitable function $C(\b)$ of $\b$.
Note that $\e_0(\b)\to 0$ for $\b\to 1$.

The map $H_\e$ is a homeomorphism of $\TTT^2$ at fixed $\Bo$. 

\* \0{\it Remark:} The result can also be interpreted as a
``shadowing theorem''. Consider the ``noisy'' trajectory $k\to\ScS_\e^k\V
x$ as a perturbation of a noiseless one. Then a noisy trajectory
starting at $\V x=(\Bo,x)$ will remain forever close to the trajectory
of the point $H^{-1}_\e(\Bo,x)$ {\it evolving under the noiseless
motion}. This is usually called a {\it shadowing property}.

\*\*
\Section(4,Heuristic considerations)
\lhead{\nota\it 4.  Heuristic considerations}

\0The notion of hyperbolicity or chaoticity can be extended,
\cite{Li01}, to random systems, for a general map $\ScS$ acting on
$\O\times \FF$, as follows: \*

\0(a) at every point $\V x=(\Bo,x)\in \O\times\FF$ the plane $W(x)$
tangent to $\FF$ contains two planes, $W^s_\Bo(x),W^u_\Bo(x)$ with
$W(x)=W^s_\Bo(x)\oplus W^u_\Bo(x)$, with positive dimensions $d^s,d^u$
and which are {\it covariant}, \ie $\ScS W^a_\Bo(x) =
W^a_{\t\Bo}(f_\Bo(x))$;

\0(b) for some $C,\k>0$:
$$\eqalign{
||\dpr_x f_{\Bo}^{*k}(x) v||&\le C e^{-k \k} ||v||,\quad v\in
W^s_\Bo(x) , \cr
||\dpr_x f^{*(-k)}_{\Bo}(x) v||&\le C e^{-k \k} ||v||,\quad v\in
W^u_\Bo(x) , \cr} \Eq(e4.1)$$
for all $k>0$; $f_{\Bo}^{*k}$ is defined below \equ(e1.4) and the
sizes $||\cdot||$ of the vectors are evaluated in the metrics at the
points to which they are applied (\ie $f_{\Bo}^{*k}(x)$ or $x$ or
${f^{*(-k)}_{\Bo}}(x)$);

\0(c) $W^a_\Bo(x)$ are continuous in $\Bo,x$.

\* 
In the example \equ(e1.3) with $\e=0$ one has $\dpr_{x}
(f_\Bo^{*(\pm k)})= S^{\pm k}_0$ so that properties (a,b,c) are satisfied,
trivially, because the map considered is an Anosov map.

It is also natural to define the expansion and contraction rates at
$(\Bo,x)$. They will simply be 
$$J_{\Bo,N}^{a}(x)=\log \Big|\Big(\det \fra{\dpr\,f_\Bo^{*(a\,N)}
(x)}{\dpr\,x}\Big)\Big|_a, \quad a=u,s , \Eq(e4.2)$$
where $(a\,N)$ means $N$ if $a=u$ and $-N$ if $a=s$
while the label $a$ appended to the Jacobian matrix means that it
is regarded as acting as a map from the plane $W^a_\Bo(x)$ to the
plane $W^a_{\t^N\Bo}(f^{*N}_\Bo(x))$.

Note that the above expansion rates are {\it not} intrinsic geometric
objects as they depend on the metric used on the manifold, as also the
constants $C,\k$ do.

The volume measure $\m_0({\rm d}x)$ on $\FF$ can be visualized, fixed
$\Bo\in\O$, in terms of the above rates: imagine to fix, for every point 
$x\in\FF$, two segments located with center in $x$
and lying on the unstable or on the stable planes through $x$, and
call them $\D^u_\Bo(x)$ and $\D^s_\Bo(x)$, respectively. Then we
can build a tiny rectangle through $x$ by considering the
line elements through $x$
$$\eqalign{&
\d_u(N,x)=f_{\t^N\Bo}^{*(-N)}\big(\D^u_{\t^N\Bo}(f_\Bo^{*N}(x))\big),
\cr &
\d_s(N,x)=f_{\t^{-N}\Bo}^{*N}\big(\D^s_{\t^{-N}\Bo}(f_{\Bo}^{*(-N)}(x))
\big) , \cr}
\Eq(e4.3)$$
and by drawing through each point $y\in \d_s$ the line element $
\d^u(N, x)$ and through each point of $y\in\d_s$ the line element $
\d^s(N, x)$ and collecting the intersections: this is well defined,
and will be called $\d_{\Bo,N}(x)$. This is close to a
rectangle with volume exponentially small as $N\to\io$ and, to
leading order in $N$, given by
$$ {\rm vol} \d_{\Bo,N}(x)= e^{-J^u_{\Bo,N}(x)} \, e^{J^s_{\Bo,N}(x)}
\g_\Bo(x) , \Eq(e4.4)$$
with, if $\a_\Bo(x)$ equals the angle between the stable and unstable
planes $W^a_{\Bo}(x)$ at $x$,
$$\g_\Bo(x)\defi \cos\a_\Bo(x) \prod_{a=u,s}
|\D^a_{\t^{(a\,N)} \Bo}(f_\Bo^{*(a\,N)}(x))| , \Eq(e4.5)$$ 
where $(a\,N)$ has the meaning explained after \equ(e4.2).
The function $\g_\Bo(x)$ is uniformly bounded away from $0$ and
$\io$. Hence the ratio of the volumes of two such parallelepipeds,
centered at $x$ and $y$, can be computed by the ratio between the
contraction factors in \equ(e4.3) evaluated at the two points
(because the basic surface elements $ \D^a_\Bo(x)$ have all the same
size and shape up to a factor $\g_\Bo(x)/\g_\Bo(y)$).

A consequence is that if we attribute to each such parallelepiped a
measure proportional to 
$$ e^{-J^u_{\Bo, 2N}(f_\Bo^{*(-N)}x)} , \Eq(e4.6) $$ 
then we expect that in the limit as $N\to\io$ the probability
distribution has a limit $\m_\Bo({\rm d}x)$ such that $\m_{\Bo}
({\rm d}x) P({\rm d}\Bo)\defi\m({\rm d}x {\rm d}\Bo)$
is an invariant distribution for the system $(\O\times\FF,\ScS_\e)$
which should describe the statistical properties of almost all data
initially chosen with the distribution $\m_0({\rm d}x) P({\rm d}\Bo)$:
\ie $\m({\rm d}x {\rm d}\Bo)$ {\it should be the SRB
distribution}. Likewise replacing in \equ(e4.6) $N$ with $-N$ and
$J^u$ with $-J^s$ one should obtain the SRB distribution for the
backwards motion (\ie for the map $\ScS^{-1}$).

\*\*
\Section(5,Overshadowing)
\lhead{\nota\it 5. Overshadowing}

\0Therefore we look for an explicit algorithm to construct a useful
representation of \equ(e4.6) or, what is the same, for the function
$J_{\Bo,1}(x)$ because
$$J^u_{\Bo, 2N}(f_\Bo^{*(-N)}x)\=\sum_{j=-N}^{N}
J^u_{\Bo,1}(f^{*j}_\Bo(x)) , \Eq(e5.1)$$
by the chain differentiation rule for composition of functions and by
the multiplication rules of determinants of matrices.

We shall use the notations of Sec.\sec(3), where the conjugation
$H_\e$, transforming a perturbation of the map
$\ScS_\e(\Bo,x)=(\t\Bo,S_0x+\e\ps_\Bo(x))$ of the torus $\O\times \TTT^2$
into a noiseless map $\ScS_0(\Bo,x)=(\t\Bo,S_0x)$, has been derived.
The homeomorphism $(\Bo',x')\otto(\Bo,x+h_\Bo(x))\=H_\e(\Bo,x)$ can be
used not only to construct the dynamics but also the stable and
unstable manifolds of each point. The latter manifolds through
$(\Bo,x+h_\Bo(x))$ are given by parametric equations of the form
$$\g_\a(t)=(\Bo,h_\Bo(x+t v_\a))\qquad t\in \RRR,
\quad \a=\pm, \Eq(e5.2)$$
where $t\to x+t\, v_\a$ is the unstable or stable manifold for
the unperturbed map $S_0$ of $\TTT^2$ into itself, respectively
if $\a=+$ or $\a=-$. 

The construction of Sec.\sec(3) gives a H\"older continuous conjugation with a
prefixed exponent $\b<1$ for a perturbation strength $\e$ that is
suitably small, depending on how close $\b$ is to $1$.  However, 
in general, the conjugation is not be differentiable: therefore
\equ(e5.2) cannot be used to compute the derivatives appearing in
the Jacobians $J^a_{\Bo,1}(x)$ and the parametrization \equ(e5.2) is
not very useful.

Instead of constructing the stable an unstable manifolds parameterized so that
the required derivatives appearing in \equ(e5.1) can be computed (or
just shown to exist) we remark that all we need are the expansion
coefficients in the stable and unstable directions of any point
$(\Bo,x)$.

To find them we apply once more the technique discussed in the
previous two sections. Calling $\widehat \Th$ the ({\it non-compact})
space $\O\times \TTT^2\times \RRR^2$ we define on it the dynamical system
$(\widehat\Th,\widehat\ScS_0)$ with
$$ \widehat\ScS_0(\Bo,x,v)=(\t\Bo, S_0 x,S_0 v) , \Eq(e5.3)$$
where $v$ has to be interpreted as a tangent vector to $\TTT^2$ in $x$
and $S_0$ is the action of the derivative of the map $x\to S_0x$ on the
tangent vector $v$.  Since the map $S_0$ is locally linear there is no
point in distinguishing between the constant matrix $\dpr_x S_0x$ and the
map $S_0$ so that we indulge in the abuse of notation implicit in
\equ(e5.3).

We can  consider the  perturbation of $\widehat\ScS_0$ defined by
$$\eqalign{
&\widehat\ScS_\e(\Bo,x,v)= \cr
&=(\t\Bo,S_0x+\e\ps_\Bo(x), S_0v+\e
( v\cdot \dpr_{x})\ps_\Bo(x)) , \cr} \Eq(e5.4)$$
where the third term is the vector $v$ tranformed by the map
$x'=x+\e\ps_\Bo(x)$ into a tangent vector 
at $x'$.

For uniformity of notation in what follows, we shall denote $S_\Bo
(x)=S_0x+\e\ps_\Bo(x)$ instead of $f_\Bo(x)$ as done so far.

This system fails to be an Anosov system not only because of the noise
(which acts trivially, however) but also because the space $\O\times
\TTT^2\times \RRR^2$ is not compact. Nevertheless we can still try to find
an isomorphism between $\widehat\ScS_\e$ and $\widehat\ScS_0$.

Success would mean that it is possible to decouple from the noise {\it
also} the evolution of the tangent vectors, \ie of the infinitesimal
displacements: not only we could match individual trajectories in the
perturbed and in the unperturbed system, as done in Sec.\sec(3), but
we could even achieve ``shadowing'' of infinitesimally close pairs of
trajectories which would split apart at the same rates as the
unperturbed ones. If possible this property could be called {\it
overshadowing}: but not surprisingly this turns out to be in general
{\it impossible} (as it will be implicit in what follows). The next
simplest goal, in trying to compare infinitesimally close pairs of
trajectories, is to try to conjugate $\widehat\ScS_\e$ with a map
$\widehat {\ScS}_{0,\e}$ defined by
$$\widehat {\ScS}_{0,\e}(\Bo,x,v)=(\t\Bo,S_0 x,
S_0v+\G_\Bo(x) v) , \Eq(e5.5)$$
with $\G_\Bo(x)$ a $2\times2$-matrix {\it commuting} with $S_0$ (hence
diagonal on the basis $v_\pm$), analytic in $\e$ and vanishing at
$\e=0$.  Therefore we look for a map $\widehat H_\e$ of a simple form
and such that $\widehat {\ScS}_\e \circ \widehat H_\e=\widehat H_\e \circ
\widehat {\ScS}_{0,\e}$, \ie
$$ \widehat H_\e\,:\,(\Bo,x,w) \otto (\Bo,x+h_{\Bo}(x)), 
w+ K_\Bo(x)w) , \Eq(e5.6)$$
where $K$ is a $2\times2$-matrix analytic in $\e$ and $0$ at $\e=0$,
to be determined. In guessing \equ(e5.6), advantage is taken
of the already known conjugation $H_\e$ between ${\ScS}_\e$ and ${\ScS_0}$,
from the analysis of Sec.\sec(3), giving us the function $h_\Bo(x)$.
\*

\0{\it Remarks:} (1) Let $\KK_\Bo(x)=1 + K_\Bo(x)$ and let
$\LL_\Bo(x)=S_0 + \G_\Bo(x)$ be functions with values in the
$2\times2$ matrices. Then the conjugation \equ(e5.6) is equivalent to
the following equation
$$ \dpr_x S_\Bo(H_\e(\Bo,x))
\KK_\Bo(x) v=\KK_{\t\Bo}(S_0 x)\LL_\Bo(x)v
\Eq(e5.7)$$ 
and to \equ(e3.3) for $H_\e$, which is an equation independent of
\equ(e5.7) (already solved in Sect.\sec(3)).  This implies that the
vector $ w_{\Bo,\pm}(x)=\KK_\Bo(x) v_\pm$ satisfies:
$$\eqalign{
&\dpr_x S_\Bo(H_\Bo(x)) w_{\Bo,\pm}(x)= \l_{\Bo,\pm}(x)
w_{\t\Bo,\pm}(S_0x)=\cr
&= (\l_\pm+\g_{\pm,\Bo}(x))w_{\t\Bo,\pm}(S_0x) , \cr}
\Eq(e5.8)
$$
\vglue1mm

\0where $\l_{\Bo,\pm}(x)$ are the diagonal elements of $\LL_{\Bo,\e}(x)$.

\0(2) The conjugation defined in \equ(e5.6) is the right one to look
at the stable and unstable directions. Its solution gives the stable
and unstable directions as functions of the unperturbed point
$H^{-1}_\e(\Bo,x)$.

\0(3) Equation \equ(e5.7) does not determine $\KK_\Bo(x)$
uniquely. Indeed, if $l_{\Bo,\pm}( x)$ are two non-zero functions
from $\TTT^2$ to $\RRR$ and if $\l_{\Bo,\pm}(x)$, $w_{\Bo,\pm}(x)$
solve \equ(e5.8), also
$$\eqalign{
&\lis \l_{\Bo,\pm}( x)={l_{\Bo,\pm}(x)\over
l_{\t\Bo,\pm}(S_0x) }\l_{\Bo,\pm}( x) , \cr&
\lis w_{\Bo,\pm}( x)=l_{\Bo,\pm}( x)
w_{\Bo,\pm}( x)\cr}\Eq(e5.9)$$ 
solve it.  To fix this ambiguity {\it we will require} that
the diagonal elements of $K_\Bo( x)$, on the basis $v_{\pm}$,
are equal to $0$, \ie the matrix $K_\Bo( x)$ is completely
off--diagonal.

\*

To simplify the rather heavy notations we denote
$$\eqalignno{
&\ps_\a(\Bo,x)\=\ps_{\Bo,\a}(x),\quad
H(\Bo,x)\=(\Bo,x+h_\Bo(x)) , \cr
&\G(\Bo,x)\=\G_\Bo(x),\quad K(\Bo,x)\=K_\Bo(x) , \cr
&\x\=\V x = (\Bo,x) , &\eq(e5.10)\cr}$$
and the equation that the matrices $K(\x),\G(\x)$ have to satisfy is a
transcription of \equ(e5.7)
$$\eqalign{
&(S_0 K(\x)-K({\ScS_0} \x) {S_0})_{ij}=\cr
&=-\e\dpr_{x_j} \ps_i( H(\x))
-\e \dpr_{x_s}\ps_i( H(\x)) K( \x)_{sj}+\cr
& +\G(\x)_{ij}+(K(\ScS_0\x)\G(\x))_{ij},\cr}\Eq(e5.11)$$
where $\dpr_{x}$ denotes a derivative of $\ps$ with respect to its
original argument and repeated indices mean implicit summation (to
abridge notations).

We write the above matrix equation on the basis in which ${S_0}$ and
$\G$ are diagonal, \ie on the basis formed by the two eigenvectors
$ v_\pm$ of ${S_0}$ in which the matrices $K_\Bo,\G_\Bo$ have been assumed
to take the form
$$ \eqalign{&
\G( \x)=\pmatrix{\g_{+}( \x)&0\cr0&\g_{-}(\x)\cr},
\cr&
K( \x)=\pmatrix{0&k_{+}(\x)\cr k_{-}(\x)&0\cr} .
\cr}
\Eq(e5.12)$$
If $\a=\pm$, $\b=-\a$ and $\dpr_{\a}\defi v_{\a}\cdot \dpr_{\Bf}$,
\equ(e5.11) becomes for $i=j$
$$\eqalign{
0=&-\e \dpr_\a \ps_{\a}(H(\x))-\e K_{\b,\a}(\x)\dpr_\b
\ps_{\a}( H(\x))+\cr
&+\g_{\a}(\x) , \cr}\Eq(e5.13)$$
and, for $i\neq j$ and if $\l_+=\l^{-1},\l_-=-\l$  are
the eigenvalues of $S_0$ (with $\l=(\sqrt5-1)/2$),
%
$$\eqalign{
(\l_\a& K_{\a,\b}(\x)-\l_\b K_{\a,\b}(\ScS_0\x)=\cr
&-\e\dpr_\b
\ps_{\a}(H(\x))-\e K_{\a,\b}(\x)\dpr_{\a}\ps_{\a}(H(x))+\cr
&+
K_{\a,\b}({\ScS_0}(\x))\g_{\b}(\x) . \cr}\Eq(e5.14)$$
%

If $\a'=\fra{\a-1}2,$ and $\a''=\fra{\a+1}2$
we can rewrite the equations \equ(e5.13) and \equ(e5.14) as
$$ \g_{\a}(\x)=\e\,\dpr_\a \ps_{\a}( H(\x))+\e\,
k_{\b}(\x)\dpr_\b \ps_{\a}( H(\x)) ,\Eq(e5.15)$$
and, respectively,
$$\eqalignno{
&k_{\a}( \x) + \l^{2} k_{\a}(\ScS_0^\a \x) 
=\a\l\,\big(-\e\dpr_\b \ps_{\a}(H(\ScS_0^{\a'}\x))-&\eq(e5.16)\cr
&
-\e k_{\a}(\ScS_0^{\a'}\x)
\dpr_\a \ps_{\a}(H(\ScS_0^{\a'}\x))
+ k_{\a}({\ScS^{\a''}_0} \x)
\g_{\b}(\ScS_0^{\a'} \x)\big) . \cr}$$
These equations are in a form
suitable for a recursive solution in powers of $\e$. For instance the
first order is
$$\eqalignno{
&\g^{(1)}_{\a}(\x)=\dpr_\a \ps_{\a}( \x) , \qquad \a=\pm , \cr
&k^{(1)}_{+}( \x) +\l^2 k_+^{(1)}({\ScS_0} \x)
= -\l\,\dpr_-\ps_+( \x) ,&\eq(e5.17) \cr
&k^{(1)}_{-}( \x)+\l^2 k_-^{(1)}({\ScS_0}^{-1} \x)
= \l\,\dpr_+\ps_{-}({\ScS_0}^{-1} \x) ,\cr}
$$
which has the solution
$$\eqalignno{
&\g^{(1)}_{\a}( \x)=\dpr_\a \ps_{\a}( \x)\qquad \a=\pm
\cr
&k^{(1)}_{+}( \x)=-\l\sum_{q=0}^\io (-1)^{q} \l^{2q}
\dpr_-\ps_{+}(\ScS_0^q \x) , 
&\eq(e5.18)\cr
&k^{(1)}_{-}( \x)= \l
\sum_{q=0}^{\io} (-1)^{q} \l^{2q} \dpr_+
\ps_{-}(\ScS_0^{-(q+1)} \x) . \cr}
$$
The equations for $\g^{(k)}_{\a}(\x)$ and $k^{(k)}_{\a}(\x)$
can be represented in graph form by suitably modifying the
similar representation derived for $ h^{(k)}_{\Bo,\a}(x)$
in Sec.\sec(3); compare Fig.2 with Fig.1.

\eqfig{184.00000pt}{56.000000pt}
{
\ins{4.48000pt}{44.000000pt}{$\st\a$}
\ins{4.48000pt}{16.800001pt}{$\st\a$}
\ins{38.40000pt}{44.000000pt}{$\st\a$}
\ins{38.40000pt}{16.800001pt}{$\st\a$}
\ins{50pt}{46.400002pt}{$\st\a\a$}
\ins{50pt}{20pt}{$\st\b\a$}
\ins{87.20000pt}{16.800001pt}{$\st\a$}
\ins{87.20000pt}{44.000000pt}{$\st\a$}
\ins{100.80000pt}{32.799999pt}{$\st\b$}
\ins{100.80000pt}{4.800000pt}{$\st\a$}
\ins{97pt}{22pt}{$\st\a\a$}
\ins{97pt}{52pt}{$\st\b\a$}
\ins{150.40000pt}{23pt}{$\st\b$}
\ins{150.40000pt}{5pt}{$\st\a$}
\ins{134pt}{17pt}{$\st\a$}
\ins{23.20000pt}{40.799999pt}{$=$}
\ins{23.20000pt}{14.400001pt}{$=$}
\ins{70.40000pt}{43.200001pt}{$+$}
\ins{70.40000pt}{15.200000pt}{$+$}
\ins{118.40000pt}{15.200000pt}{$+$}
\ins{12.80000pt}{36.000000pt}{$\st (k)$}
\ins{12.80000pt}{6.400000pt}{$\st (k)$}
\ins{56.79999pt}{6.400000pt}{$\st (k-1)$}
\ins{56.79999pt}{36.000000pt}{$\st (k-1)$}
\ins{116.80000pt}{51.200001pt}{$\st (p)$}
\ins{116.80000pt}{33.600002pt}{$\st (k-1-p)$}
\ins{116.80000pt}{23.200001pt}{$\st (p)$}
\ins{116.80000pt}{4.480000pt}{$\st (k-1-p)$}
\ins{164.80000pt}{24.000000pt}{$\st (p)$}
\ins{164.80000pt}{7.200000pt}{$\st (k-p)$}
}
{fig2}{}

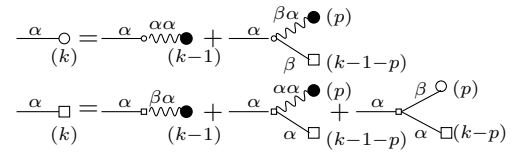
\* \0Fig.2 {\nota Here $\a=\pm$ and $\b=-\a$.  All the
lines have to imagined to carry arrows (not drawn) pointing toward the
root.  The line carrying a label $\a$ and emerging from a circle or a
square with label $(k)$ denotes $\g_{\a}^{(k)}$ or
$k_{\a}^{(k)}$, respectively.  The wavy line emerging from a
bullet with label $(p)$, with $1\le p< k-1$, ending in a small circle
and carrying a pair of labels $\g,\d$, represents $[\dpr_{\g}
\ps_{\d}]^{(p)}(H(\x))$, the $p$-th order in the power expansion in
$\e$ of $(\dpr_{\g} \ps_{\d})(H(\x))$.  The small square in the
node closest to the root into which a wavy line arrives will carry a
label $q=0,1,2\ldots $ (see the first two graphs in the second line)
and it expresses that $H(\ScS_0^{\a'+ \a \tilde q_\a}(\x))$ (in the
second line), with $\tilde q_+ = q$ and $\tilde q_-=q+1$, is
the argument in which the functions $\dpr_{\g} \ps_{\d}$ are
computed.  Furthermore a summation over
$q=0,1,\ldots$ and a multiplication by $-\a(-1)^{q}\l^{1+2q}$ is
understood to be performed over the nodes represented as small
squares. The last graph in the second line does not carry any further
labels and it represents the last term in \equ(e5.16).\vfil} \*

The representation is drawn in figure Fig.2 and the symbols are
explained in the corresponding caption: the reader will recognize in
them a pictorial rewriting of \equ(e5.16).

In this case too we can continue the expansion until in the r.h.s. of
Fig.2 all top nodes of the graph are either squares or circles
carrying a label $(1)$, \ie they represent a first order contribution
to $\G$ (circle) or to $K$ (square), or bullets representing either
$\dpr_\a \ps_{\b}(\x)$ or $\dpr_\a \ps_{\b}(\ScS_0^{\a'+\a
\tilde q_{\a}} \x)$; cf. the caption of Fig.2 for notations.
Of course the latter quantities can themselves be represented by the
tree expansion discussed in Sec.\sec(3): if we do so then we obtain a
full expansion in powers of $\e$ in which the wavy lines with label
$p$ are replaced by a tree with $p$ nodes.

The rule to construct the value of each tree graph is easily read from
\equ(e5.11) and from the rules discussed above, and in
Sec.\sec(3), to build the value of trees representing $ h$.

The estimate of the $k$--th order contribution is given by
\equ(e3.14) with an extra factor $N^k$ to take into account the
extra derivatives due to the lines with two labels. Also the counting
of the trees has to be modified but the end result will be that $\G$
and $K$ are expressed by convergent series in $\e$ for $|\e|<\e_0(\b)$,
where $\e_0(\b)$ can be taken of the form \equ(e3.15) with a different
numerical factor and with $N$ replaced by $N^{2}$.

\*\*
\Section(6,Construction of the SRB distribution)
\lhead{\nota\it 6. Construction of the SRB distribution}

\0Consider the partition $\EE$ of $\O\times\FF$ by the sets
$E_{\o,\s}=C_\o\times P_\s$ where $\CC=\{C_1,\ldots,C_m\}$ is the
partition of $\O$ according to the value of the symbol $\o_0$ and
$\PP_0=\{P_1,\ldots,P_n\}$ is a Markovian pavement of $\FF$ for the
map $S_0$, \cite{GBG04}. It is convenient to suppose that the
Markovian pavement $\PP_0$ is ``reversible''. This means that
there exists a map $I_0$, with the property that $I_0^2 = \pm 1$ and
$I_0\crcl S_0^k = S_0^{-k}\crcl I_0$ for some integer $k$, such that
for all $\s=1,\ldots,n$ one has $I_0P_\s=P_{\s^T}$ for some $\s^T$
(this is not restrictive as it can be achieved by considering
the new pavement $I_0\PP_0\cap\PP_0$). Given a point $\x$ let
$(\Bo,\Bs)$ be its ``history'' on the partition $\EE$ for the map
$\ScS_0$. This means that $\ScS_0^k\x\in E_{\o_k,\s_k}$ for all $k$.
In terms of the sequences $(\Bo,\Bs)$ the motion is simply
a translation $(\t,\r)$ such that $(\t \Bo,\r\Bs)_j=
(\Bo,\Bs)_{j+1}$, \cite{GBG04}.

The partition $\EE'=H_{\e}(\EE)$ is then Markovian for the
perturbed system. A correspondence can be generated between points in
$\O\times\FF$ by associating two points $\x'$ and $\x$ if the first has
history $(\Bo,\Bs)$ on $\EE'$ identical to the history of $\x$ on the
partition $\EE$ under the action of the map $\ScS_0$: we denote
$\x'=X_\e(\Bo,\Bs)$ and $\x=X_0(\Bo,\Bs)$. The correspondence so defined
is identical to the correspondence between $\x'$ and $\x$ defined by
$\x'=H(\x)$. And in Sec.\sec(5) it has been proved that

\*

{\it There exists $\e_{0}(\b)$ such that the expansion rate
$\l_{u,\Bo}(\Bs)$ of ${\ScS}_\e$ along the unstable manifold of $ \x'$
is defined and holomorphic in $\e$ in the disk $|\e|<\e_0(\b)$. As a
function of $(\Bo,\Bs)$ it is H\"older continuous with some
exponent $0<\b<1$ and modulus $C(\b)$.}
\*

\0This can be seen as follows: one just notes that if $ \x'=H_\e(\x)$
then the unstable direction at $ \x'$, which is a vector tangent to
$\FF$, will be ${w}_{\Bo,+}(x)= {v}_++K_\Bo( x){ v}_+$.  Then the
expansion rate along the unstable manifold of $\x'=(\Bo,x')=H_\e(\x)$
will be $e^{A_{\Bo,u}(\Bs)}\defi |J^u_{\Bo,1}(x')|$ and, by
\equ(e5.8),\equ(e4.1),\equ(e4.2),
$$A_{\Bo,u}(\Bs)=
\log \Big(\big(\lambda_++\g_{\Bo,+}(x)\big)
\fra{|{w}_{\Bo,+}(S_0 x)|}{|{ w}_{\Bo,+}( x)|}\Big) ,
\Eq(e6.1)$$
where $|{w}_{\Bo,+}( x)|=\sqrt{1+k_{\Bo,+}(x)^2}$.

The functions $\g_{\Bo,+},k_{\Bo,+}$ are analytic in $\e$ and H\"older
continuous in $\x$ with a uniformly bounded modulus $C(\b)$ if $\b<1$
and $\e$ is small enough. Since $\{(\o_i,\s_i)\}_{i=-\io}^\io$ fixed
means $\x$ fixed (because $\Bs$ is the history of a point $ x$ on a
Markov pavement for the $\e$--independent map ${S}_0$), we see that
$A_{\Bo,u}(\Bs)$ is analytic in $\e$ at fixed $(\Bo,\Bs)$ and H\"older
continuous in $(\Bo,\Bs)$ and therefore in $ \x$. 

Hence the function $A_{\Bo,u}(\Bs)$ generates a short range potential
on the one--dimensional ``spin system'' whose states are the sequences
$(\Bo,\Bs)$, in the sense of \cite{Ma99}.
\*

An identical argument holds for the similar function
$A_{\Bo,s}(\Bs)\defi\log |J_{\Bo,1}^s(x')|$. It follows, by standard
arguments on one--dimensional Gibbs distributions, see Chap.6 in
\cite{GBG04}, that the ``short range'' Gibbs distribution $\n$ with
formal energy function
$$\sum_{k=-\io}^0 A_{\t^{k}\Bo,s}(\r^{k}\Bs)+\sum_{k=0}^\io
A_{\t^{k}\Bo,u}(\r^{k}\Bs) 
\Eq(e6.2)$$
is {\it well defined} and it has the property that if $C^N_{\Bo,\Bs}$
is defined as $\{\Bo',\Bs'\,|\, \o'_i=\o_i,\,\s'_i=\s_i,\,\forall
|i|\le N\}$, then
$$B^{-1}\le \fra{\n(C^N_{\Bo,\Bs})}{{\rm
vol}(C^N_{\Bo,\Bs})}\le B\Eq(e6.3)$$
for a suitable constant $B>0$.

Therefore the distribution $\n$ is absolutely continuous with respect
to the volume distribution (since $\g_\Bo$ in \equ(e4.6) is bounded
above and below).  

The distribution $\n$ is {\it not} translation--invariant because its
Gibbs potential in the ``far future'' is governed by $A_{\Bo,u}(\Bs)$
while in the far past it is governed by $A_{\Bo,s}(\Bs)$, see
\equ(e6.2). But short range Gibbs distributions (whether invariant or
not) enjoy exponentially mixing properties: hence the distribution
$\n$, and consequently the volume distribution, will be very close on
the evolution of the cylinders, for very large times,
to the invariant Gibbs distributions with (formal) potentials
$$ 
\sum_{k=-\io}^{+\io} A_{\t^{k}\Bo,a}(\r^k\Bs), \quad
\cases{{\rm as}\
  p\to+\io\ {\rm if}\ a=u , &\cr
{\rm as}\  p\to-\io\ {\rm if}\ a=s ,
\cr}\Eq(e6.4)$$
respectively.  Therefore the distribution with potential in
\equ(e6.4) with $a=u$ is the SRB distribution for the forward
evolution while for $a=s$ the distribution is the SRB distribution for
the backward evolution.

\*\*
\Section(7,Time reversal symmetry)

\0The result of Sec.\sec(6) allows us to conclude that the SRB
distribution $\m_{\rm srb}$ will be a ``space-time'' Gibbs state for the
energy function $A_{\Bo,u}(\Bs)$ in the sense of \cite{Ma99}.

\lhead{\nota\it 7. Time reversal symmetry}

Time reversal symmetry is a symmetry that should be inherited from the
microscopic time reversal symmetry of purely Hamiltonian dynamics. It
may be destroyed by the thermostats, or better, by the models that are
introduced to describe them.

In presence of noise the matter is even more involved. However under
the assumptions that we have posed upon $\ScS_\e$ it is possible to
show that if the noiseless system $\ScS_0$ is reversible a ``kind'' of
time reversal symmetry will continue to hold when the noise is not
neglected provided the noise also has a reversibility property.

We supposed, since the very beginning in Sec.\sec(1), reversibility
of the stationary probability distribution $P$ on $\O$ in the sense
that the operation $\th$ defined by $\th\Bo=\Bo^T$ with
$\o^T_i=\o_{-i}$ (``history reversal'') conserves the distribution
$P$:
$$P(E^T)=P(E), \Eq(e7.1)$$
where $E^T=\{\Bo^T : \Bo \in E\}$, for all (Borel-)sets $E$.

Suppose that the unperturbed system evolution on $\FF$, that we
assume to be an Anosov map $f$, is reversible in the sense that
there is a map $I_0:\FF\to\FF$  with the property $I_0 \crcl
f^k=f^{-k} \crcl I_0$ for some $k$ and $I_0^2 = \pm 1$,
or more generally $I_0^{k'}=1$ for some $k'$ even.

The above assumption is satisfied in the examples treated in the
previous sections because the time reversal operation
$I_0(\f_1,\f_2)=(\f_2,-\f_1)$ has the property $I_0 \crcl S^2_0=
S_0^{-2} \crcl I_0$, if $S_0$ is given by \equ(e1.1).

The problem is that it is not clear that a time reversal symmetry
holds when there is a perturbation, even if small.  But, if $\e$ is
small, the time reversal symmetry $\II_0(\Bo,x) =(\Bo^T,I_0x)$ is not
``broken'' by the perturbation and the noisy system admits a time
reversal symmetry $\II_\e$. The latter can be immediately written as
$$\II_\e \defi H_\e \crcl \II_0 \crcl H_\e^{-1} . \Eq(e7.2)$$
{\it However} for the same reasons for which the conjugacy between
$\ScS_\e$ and $\ScS_0$ was not smoother than H\"older continuous it
will turn out that $\II_\e$ is also not necessarily smoother than
H\"older continuous.

In terms of symbolic dynamic the time reversal $\II_\e$ acts on a
point whose symbolic history is $(\Bo,\Bs)$ by transforming it into
the point with symbolic history $(\Bo^T,\s^T)$, where
$(\Bs^T)_i=\s^T_{-i}$, if time reversal
transforms the element $P_\s$ of the Markov pavement for $S_0$ into
$P_{\s^T}$. It is therefore clear that $\II_\e^{k'}=1$
if $I_0^{k'}=1$.

The lack of smoothness of the new time reversal implies that the
fluctuation theorem can certainly be proved only in cases in which the
perturbed system also admits a smooth time reversal. Then the proof is
the {\it same} as the one for Anosov maps because the latter was based
on the existence of a Markovian partition of phase space and on the
Gibbs property of the SRB distribution with a Gibbs weight generated
by a function $A_{u,\Bo}(\Bs)$, as shown in Sec.\sec(6), satisfying
$A_{u,\Bo^T}(\Bs^T)=- A_{s,\Bo}(\Bs) $ (which follows from time
reversal symmetry and from the isometric nature of such reversal)
\cite{GC95,Ga95b}.

Examples can be obtained easily: consider the paradigmatic example
$\ScS_\e$ of the previous sections and the map $\TT(\Bo,x,\Bh,y)$
defined on $\O\times \TTT^2\times\O\times \TTT^2$ by
$$\TT(\Bo,x,\Bh,y)=(\ScS_\e(\Bo,x),\ScS_\e^{-1}(\Bh,y)) . \Eq(e7.3)$$
Then a time reversal symmetry (\ie an isometry anticommuting with
time and squaring to the identity) can be defined as
$\II((\Bo,\Bh,x,y)=(\Bh,\Bo,y,x)$ and the fluctuation theorem
applies to this system.

The last example is somewhat artificial: in applications
time reversal should be a built-in symmetry so that its checking
should be immediate. This is often the case in non--random systems.
Note that in numerical simulations realizing the system $\TT$ out of a
system $\ScS$ would be easy. 

\*\*
\Section(8,Structural stability)
\lhead{\nota\it 8. Structural stability}

\0All what has been said about the existence of the conjugacies to the
unperturbed system can be extended to the cases in which the torus
has arbitrary dimension and $S_0$ is an hyperbolic linear torsion (\ie
it is a map generated by a matrix $S_0$ with determinant $\pm1$, integer
entries and all eigenvalues with modulus different from $1$), by
replacing $\l_\pm$ by the determinants of $S_0$ restricted to
the stable or unstable planes.

Consider the case in which the map $x\to f(x)$ is just an Anosov map
of a general manifold $\FF$, still assuming analytic regularity.  If the
unperturbed Anosov map $f$ is close to some hyperbolic linear torsion
$S_0$ of a torus it will be possible to find an analytic conjugacy to
the systems to $\ScS_0$ simply because we can think that the
perturbation $\e\ps_\Bo$ contains a non-random part which is the
perturbation of $S_0$ into $f$. The same can be said of maps that can
be analytically conjugated to hyperbolic torsions.

To extend the results on the fluctuation relation, however, the strong
assumption about time reversal has to be always added.

More interesting and subtle is the further question of how much the
above analysis depends on the assumption that we are perturbing a
system close to one $\ScS_0$ in which a reversible noise and a linear
torsion of a torus $S_0$ do not interact. 

Under the only assumption that $S_0$ is analytic and Anosov the
equations for $H$ will have to be written in coordinates and possibly
several coordinate charts will have to be used. An apparently
unsurmountable difficulty may seem that the directions $v_\pm$ will
now depend on $x$ and therefore the solution algorithm will simply
fail. A more attentive study of the equations shows, nevertheless,
that the equations are {\it soluble} order by order as at every order
the curvature of the manifolds systematically gives contributions of
higher order, see \cite{BFG04} for details showing that the existence
of a convergent power series in $\e$ for $H$ exists. Since that is the
main difficulty there is little doubt that the analysis that we have
performed in detail for the case of the random perturbations of
\equ(e1.1) carries unchanged to the last case.

\*

\bibliographystyle{apsrev}

\revtex\ \&\ \EqaligN \ \&\ \fancyhdr

\end{document}